\documentclass[submission, Proceedings]{SciPost}

\hypersetup{
    colorlinks,
    linkcolor={red!50!black},
    citecolor={blue!50!black},
    urlcolor={blue!80!black}
}

\usepackage[bitstream-charter]{mathdesign}
\DeclareSymbolFont{usualmathcal}{OMS}{cmsy}{m}{n}
\DeclareSymbolFontAlphabet{\mathcal}{usualmathcal}

\usepackage{color}
\usepackage{bm}
\usepackage{amsmath}
\usepackage{booktabs}
\usepackage{float}
\usepackage[caption=false]{subfig} 
\usepackage{tabularx}
\usepackage{scalerel}
\usepackage{tikz}
\usetikzlibrary{svg.path}

\usepackage{hyperref}

\newcommand{\bfa}{\boldsymbol{a}}
\newcommand{\bfy}{\boldsymbol{y}}

\newcommand{\bfk}{\boldsymbol{k}}

\usepackage[normalem]{ulem}

\newcommand*\oline[1]{%
   \vbox{%
     \hrule height 0.5pt
     \kern0.4ex
     \hbox{%
       \kern-0.15em
       \ifmmode#1\else\ensuremath{#1}\fi
       \kern-0.15em
     }
   }
}

\begin{document}

\begin{center}{\Large \textbf{
Reweighting the quark Sivers function with STAR jet data
}}\end{center}

\begin{center}
C.~Flore\textsuperscript{1$\star$},
M.~Boglione\textsuperscript{2,3},
U.~D'Alesio\textsuperscript{4,5},
J.~O.~Gonzalez-Hernandez\textsuperscript{2,3},
F.~Murgia\textsuperscript{5} and
A.~Prokudin\textsuperscript{6,7}
\end{center}

\begin{center}
{\bf 1} Universit\'e Paris-Saclay, CNRS, IJCLab, 91405 Orsay, France
\\
{\bf 2} Dipartimento di Fisica Teorica, Universit\`a di Torino, Via P.~Giuria 1, Torino, I-10125, Italy
\\
{\bf 3} INFN, Sezione di Torino, Via P.~Giuria 1, Torino, I-10125, Italy
\\
{\bf 4} Dipartimento di Fisica, Universit\`a di Cagliari, Cittadella Universitaria, I-09042 Monserrato (CA), Italy
\\
{\bf 5} INFN, Sezione di Cagliari, Cittadella Universitaria, I-09042 Monserrato (CA), Italy
\\
{\bf 6} Division of Science, Penn State University Berks, Reading, Pennsylvania 19610, USA
\\
{\bf 7 }Theory Center, Jefferson Lab, 12000 Jefferson Avenue, Newport News, Virginia 23606, USA

* carlo.flore@ijclab.in2p3.fr
\end{center}

\begin{center}
\today
\end{center}


\definecolor{palegray}{gray}{0.95}
\begin{center}
\colorbox{palegray}{
  \begin{tabular}{rr}
  \begin{minipage}{0.1\textwidth}
    \includegraphics[width=22mm]{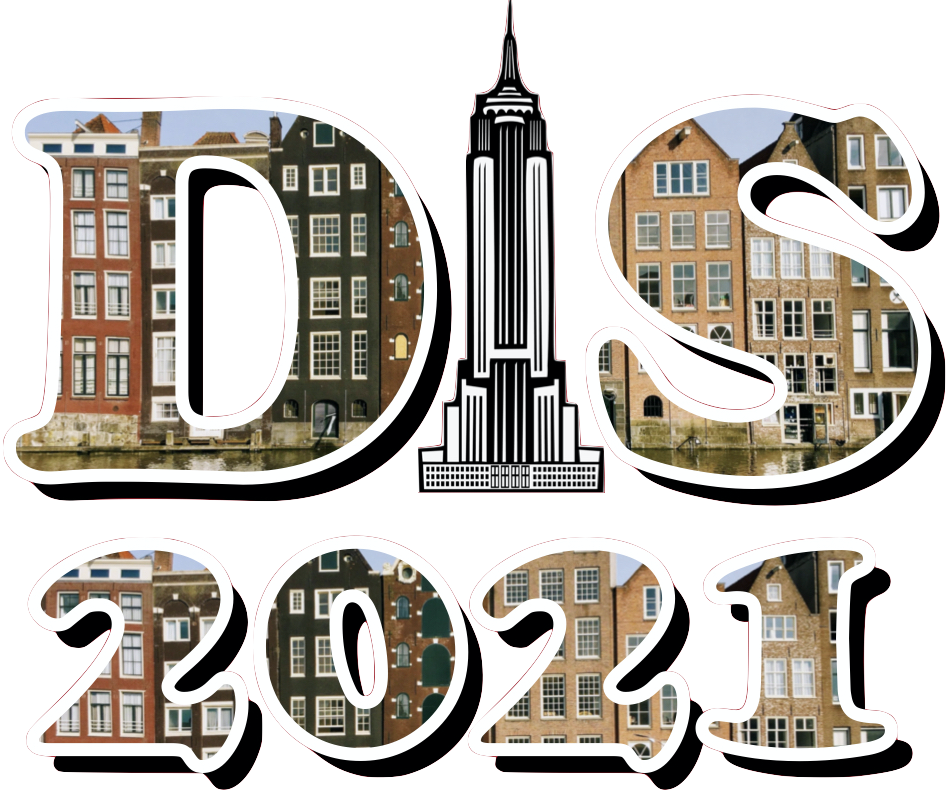}
  \end{minipage}
  &
  \begin{minipage}{0.75\textwidth}
    \begin{center}
    {\it Proceedings for the XXVIII International Workshop\\ on Deep-Inelastic Scattering and
Related Subjects,}\\
    {\it Stony Brook University, New York, USA, 12-16 April 2021} \\
    \doi{10.21468/SciPostPhysProc.?}\\
    \end{center}
  \end{minipage}
\end{tabular}
}
\end{center}

\section*{Abstract}
{\bf
The Bayesian reweighting procedure is applied for the first time to a TMD distribution, the quark Sivers function extracted from SIDIS data. By exploiting the recent published single spin asymmetry data for the inclusive jet production in $p^\uparrow p$ collisions from the STAR collaboration at RHIC, we show how such a procedure allows to incorporate the information contained in the new data set, without the need of re-fitting, and to explore a much wider $x$ region compared to SIDIS measurements. The reweighting method is also extended to the case of asymmetric errors, and the results show a significant improvement on the knowledge of the quark Sivers function.
}

\section{Introduction}
\label{sec:intro}

The three-dimensional structure of nucleons can be described in terms of Transverse Momentum Dependent (TMD) quark and gluon distributions. At leading twist, among the eight independent quark TMDs, the Sivers function $f_{1T}^\perp$~\cite{Sivers:1989cc,Sivers:1990fh} is one of the most studied and plays a seminal role. It is a genuine TMD distribution that encodes the correlation between the transverse polarization of the nucleon and the intrinsic transverse momentum of the quarks inside the nucleon. At variance with unpolarized TMD PDFs, it is also expected to be process dependent, changing sign when probed in Semi-Inclusive DIS (SIDIS) and Drell-Yan (DY) processes~\cite{Collins:2002kn, Brodsky:2002rv}. A non-zero Sivers function is also an indirect signal of nonvanishing parton orbital angular momentum. 

Here, we report on the findings of Ref.~\cite{Boglione:2021aha}, where we applied, for the first time, a reweighting procedure to a TMD density.

\section{Formalism}

The quark Sivers function is usually extracted from the SIDIS azimuthal asymmetries $\,A_{UT}^{\sin(\phi_h - \phi_S)}$ $\equiv F_{UT}^{\sin(\phi_h - \phi_S)} / F_{UU} = \mathcal{C}\left[f_{1T}^{\perp q} D_1^q\right] / \mathcal{C} \left[f_1^q D_1^q\right]$. At the same time, its corresponding effect could be responsible for the transverse single-spin asymmetries (SSAs) measured in $p^\uparrow p \to {\rm jet}\;X$ processes. At variance with SIDIS processes, for which we detect two separate energy scales ($Q_2 \gg Q_1$ $\sim \Lambda_{\rm QCD}$), in this latter class of reactions only a single, hard scale is measured. Although in principle such single-scale processes are described within the collinear twist-3 approach~\cite{Gamberg:2013kla}, one can also use some alternative, phenomenological approaches such as the generalized parton model (GPM)~\cite{DAlesio:2004eso,Anselmino:2005sh,DAlesio:2007bjf} and its color gauge invariant version (CGI-GPM)~\cite{Gamberg:2010tj,DAlesio:2011kkm, DAlesio:2017rzj,DAlesio:2018rnv}. Within these effective models, a factorized formulation in terms of TMDs is assumed as a starting point. In the GPM, the Sivers function is considered to be the same as extracted in SIDIS measurements, and no sign-change effect is taken into account. The sign change is recovered in the CGI-GPM, by including initial and final state interactions within a one-gluon exchange approximation. In the spirit of testing the compatibility of the extraction of the Sivers function from SIDIS data, we analyzed the recent SSA data for inclusive jet production in $pp$ collisions from the STAR Collaboration at RHIC~\cite{Adam:2020edg}, within the GPM and CGI-GPM approaches. 

The SSA for inclusive jet production in polarized $pp$ collisions is defined as 
\begin{equation}
A_N\equiv \frac{d\sigma^\uparrow - d\sigma^\downarrow}{d\sigma^\uparrow + d\sigma^\downarrow} \equiv \frac{d\Delta\sigma}{2d\sigma}. 
\end{equation}
In the CGI-GPM approach, numerator and denominator of the asymmetry are given by~\cite{Gamberg:2010tj}:
\begin{equation}
\label{eq:sivgen}
\begin{aligned}
 d\Delta\sigma^{\rm CGI-GPM}\, = \frac{2\alpha_s^2}{s}\sum_{a,b,c,d}& \int \frac{dx_a \, dx_b}{ x_a \, x_b} \; d^2\bm{k}_{\perp a} \, d^2\bm{k}_{\perp b} \left ( -\frac{k_{\perp a}}{M_p} \right  ) f^{\perp a}_{1 T}(x_a, k_{\perp a})\cos\varphi_a\\
 &\times  f_{b/p}(x_b, k_{\perp b})\,H^{\rm Inc}_{ab \to cd}
\> \delta(\hat s + \hat t + \hat u) \>,
\end{aligned}
\end{equation}
\begin{equation}
\label{eq:unp}
 d\sigma\,= \frac{\alpha_s^2}{s}\sum_{a,b,c,d} \int \frac{dx_a \, dx_b}{ x_a \, x_b} \; d^2\bm{k}_{\perp a} \, d^2\bm{k}_{\perp b}
f_{a/p}(x_a, k_{\perp a}) 
   f_{b/p}(x_b, k_{\perp b})\,H^U_{ab \to cd}
\> \delta(\hat s + \hat t + \hat u) \>,
\end{equation}
where $\alpha_s$ is the strong coupling constant, $s$ is the $pp$ center-of-mass energy, and  $\hat s$, $\hat t $, $\hat u$ are the usual Mandelstam variables for the partonic subprocess $ab\to cd$. Moreover, $f_{b/p}(x_b, k_{\perp b})$ is the unpolarized TMD distribution for parton $b$. Notice that in a leading-order approach the jet is identified with the final parton $c$. Finally, $H^{\rm Inc}_{ab \to cd}$'s are the perturbatively calculable hard scattering functions, that can be found in Ref.~\cite{Gamberg:2010tj} for the case when $a = q, \bar{q}$. The GPM expressions are obtained from Eq.~(\ref{eq:sivgen}) by simply replacing $H^{\rm Inc}_{ab \to cd}$ with the standard unpolarized partonic cross sections, $H^U_{ab \to cd}$.

\subsection{The reweighting procedure}
\label{sec:reweighting}

In order to assess the impact of the new STAR data on the extraction of the Sivers function, we adopt a reweighting procedure. Such a technique has been already used in the context of usual collinear PDFs~\cite{Giele:1998gw,Ball:2010gb,Sato:2013ika,DAlesio:2020vtw}, but so far it has never been applied to a TMD density. 

In brief, the reweighting procedure works as follows. Let us consider a model for a TMD depending on a set of parameters $\bfa = \{ a_1, \cdots, a_n\}$ with prior probability distribution $\pi(\bfa)$. Defining the $\chi^2$ for a specific set of data $\bfy$ as:
\begin{equation}\label{eq:chi2}
    \chi^2[\bfa, \bfy] = \sum_{i,j = 1}^{N_{\rm dat}} (y_i[\bfa] - y_i)\, C_{ij}^{-1}(y_j[\bfa] - y_j)\,,
\end{equation}
one finds the best fit $\bfa_0$, by usual $\chi^2$ minimization, that renders a minimum value $\chi^2_0$. The uncertainty on the extracted TMD is then calculated by generating $k = 1,\cdots, N_{\rm set}\,$ Monte Carlo (MC) sets $\bfa_k$\footnote{In Ref.~\cite{Boglione:2021aha}, we generated $N_{\rm set} = 2 \cdot 10^5$ MC sets adopting a Markov-Chain MC procedure with Metropolis-Hastings algorithm \cite{10.2307/2684568}.}. Each of these sets have a corresponding $\chi^2_k[\bfa_k,\bfy]$ (calculable using Eq.~(\ref{eq:chi2})) within a certain tolerance: $\chi^2_k \in [\chi^2_0: \chi^2_0 + \Delta\chi^2]$. By using Bayes theorem, one calculates the posterior density given the data:
\begin{equation}
   {\cal P}(\bfa| \bfy )=\frac{{\cal L} (\bfy | \bfa)\, \pi(\bfa)}{\mathit{Z}}\; ,
   \label{eq:posterior}
\end{equation}
where ${\cal L} (\bfy | \bfa)$ is the likelihood and $\mathit{Z} \equiv {\cal P}(\bfy)$ is the evidence. Following Refs.~\cite{Giele:1998gw,Sato:2013ika,DAlesio:2020vtw}, we adopt an exponential form of the likelihood, with weights:
\begin{equation}\label{eq:weights}
 w_k (\chi^2_k) = \frac{{\rm exp}\left\{-\frac 12\,\chi^2_k[\bfa_k,\bfy]\right\}}{\sum\limits_i w_i}
\end{equation}
that can be used to calculate expectation values and variances of an observable ${\cal O}[\bfa_k]$ as ${\rm E}[{\mathcal O}]$ $\simeq \sum_k w_k \,{\mathcal O}(\bfa_k)$, $\;{\rm V}[{\mathcal O}] \simeq \sum_k w_k \left({\mathcal O}(\bfa_k) - {\rm E}[{\mathcal O}]\right) ^2$ respectively. Such quadratic forms render only symmetric uncertainties, and to properly take into account non Gaussian distributions, we extend this method calculating asymmetric uncertainties. In what follows, the median is used at central value, and the asymmetric errors are given at $2\sigma$ confidence level (CL).

New data $\bfy_{\rm new}$ will change the weights $w_k \to w_k^{\rm new} \left(\chi^2_k + \chi^2_{{\rm new}, k}\right)$ and so the posterior densities will vary, indicating the impact of such new data on the extracted TMD.

\section{Results}

We apply the Bayesian reweighting procedure of Section~\ref{sec:reweighting} to the following quark Sivers function parametrization, extracted in Ref.~\cite{Boglione:2018dqd} from $N_{\rm dat}^{\rm SIDIS} = 220$ datapoints\footnote{The corresponding $\Delta\chi^2$ for $N = 5$ parameters at $2\sigma$ CL is $\Delta\chi^2 = 11.31$.}: 

\begin{equation}\label{eq:Sivers-parametrization}
    \Delta^N\!f_{q/p^\uparrow}(x,k_\perp) = \frac{4 M_p k_\perp}{\langle k_\perp^2\rangle_S} \Delta^N\!f_{q/p^\uparrow}^{(1)}(x) \frac{e^{-k_\perp^2/\langle k_\perp^2\rangle_S}}{\pi \langle k_\perp^2\rangle_S}\,.
\end{equation}
Here, $q = u,\,d$, and $\Delta^N\!f_{q/p^\uparrow}^{(1)}(x)$ is the Sivers first $k_\perp$-moment:
\begin{equation}\label{eq:f1Tp-first-mom}
    \Delta^N\!f_{q/p^\uparrow}^{(1)}(x) = \int d^2 \bfk_\perp \frac{k_\perp}{4 M_p} \Delta^N\!f_{q/p^\uparrow}(x,k_\perp) \equiv - f_{1T}^{\perp (1) q}(x) = N_q\,(1-x)^{\beta_q}\,.
\end{equation}

As new evidence, we consider the recent STAR data~\cite{Adam:2020edg}, that have a wide coverage in $x_F$ $= 2 P_L / \sqrt{s} \in [0.1:0.6]$. We stress that such a region is complementary to SIDIS measurements, and can give important information on the poorly constrained large-$x$ behavior of the Sivers function. Notice also that, as these data are referred to electromagnetic jets, we select the subset of data with photon multiplicity $n_\gamma > 2$, as it is not contaminated by single photon or $\pi^0$ production contributions. 

\begin{figure}[h!]
\centering
\includegraphics[width=.45\textwidth]{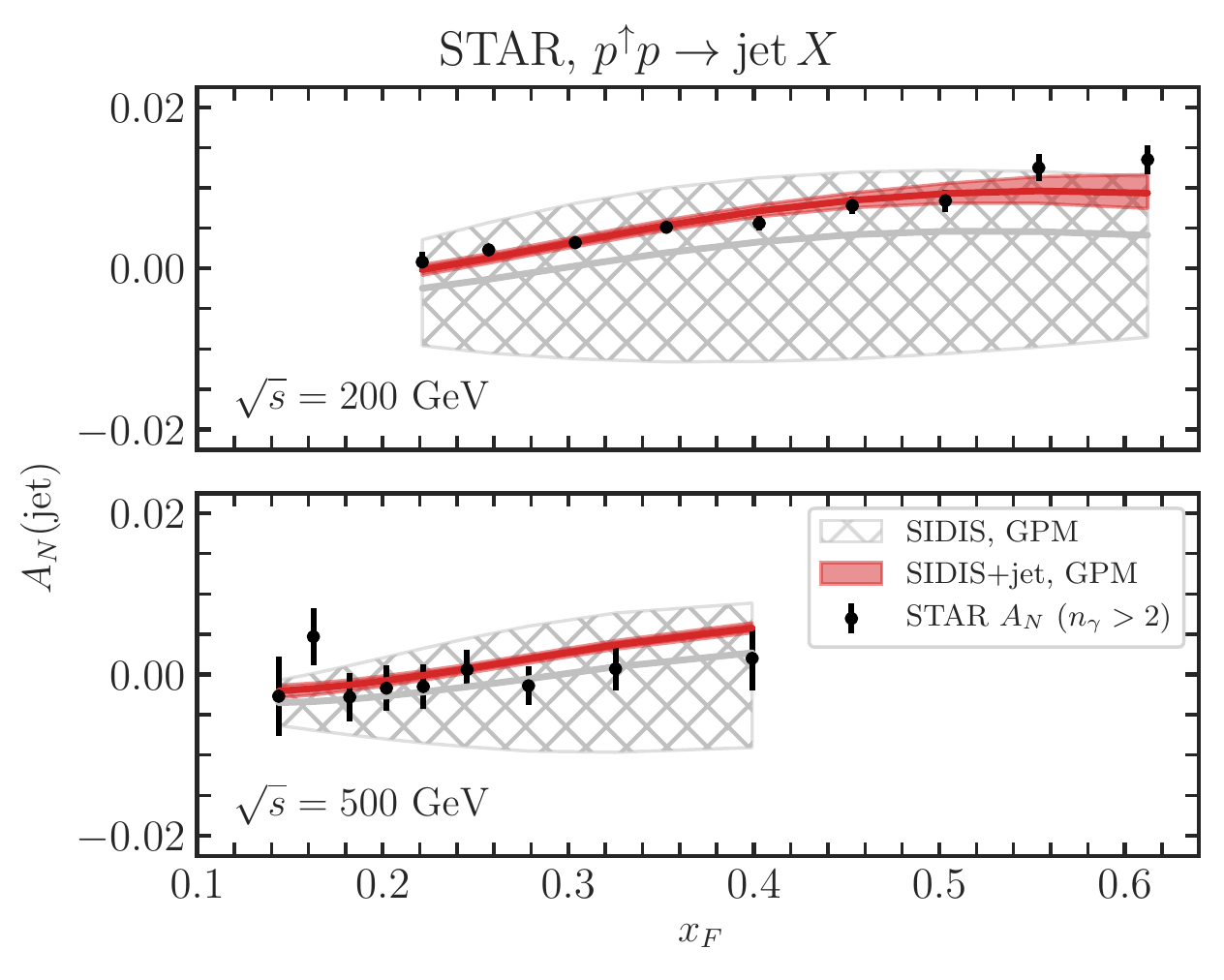}
\includegraphics[width=.45\textwidth]{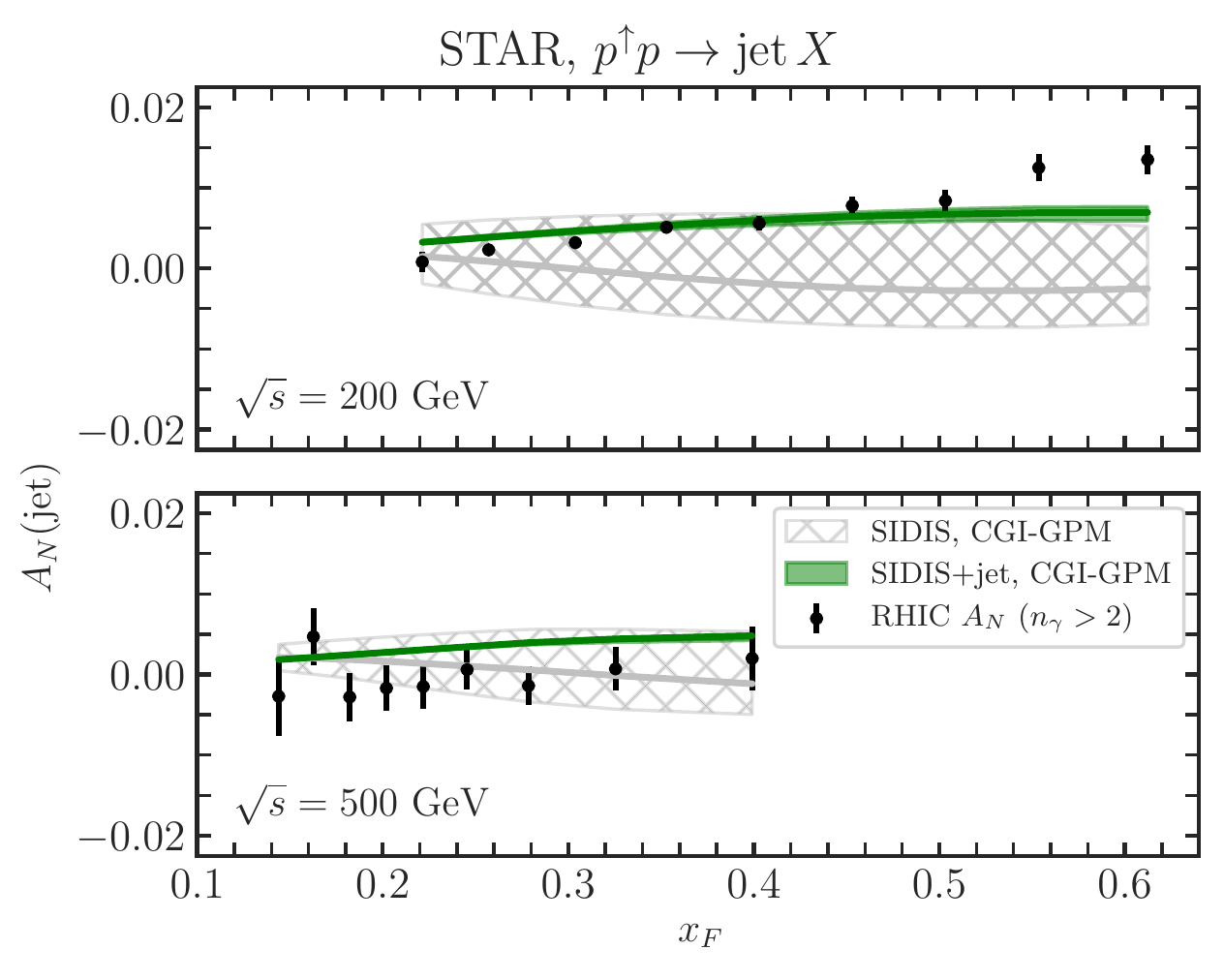}
\caption{Results for the reweighting procedure from SIDIS and $A_N$ jet data in the GPM (left) and CGI (right) formalisms, compared with STAR measurements~\cite{Adam:2020edg} at $\sqrt{s} = 200\,\text{GeV}$ (upper panels) and $\sqrt{s} = 500\,\text{GeV}$ (lower panels). Uncertainty bands are at $2\sigma$ CL. The results before (hatched grey bands) and after (solid red/green bands) reweighting are shown.}
\label{fig:reweigthing-SIDIS+jet}
\end{figure}

In Fig.~\ref{fig:reweigthing-SIDIS+jet} we show the results of the reweighting procedure for the $A_N$ predictions at STAR kinematics in the GPM (left, red) and the CGI-GPM (right, green) approaches. The grey hatched curves and bands are relative to the predictions based on SIDIS data only, while the solid colorful ones are the reweighted curves, dubbed as ``SIDIS+jet''. Although the predictions from SIDIS already describe the data within large uncertainties, the reweighted curves show a good improvement and reduced errors.

\begin{figure}[h!]
\centering
\includegraphics[width=.55\textwidth]{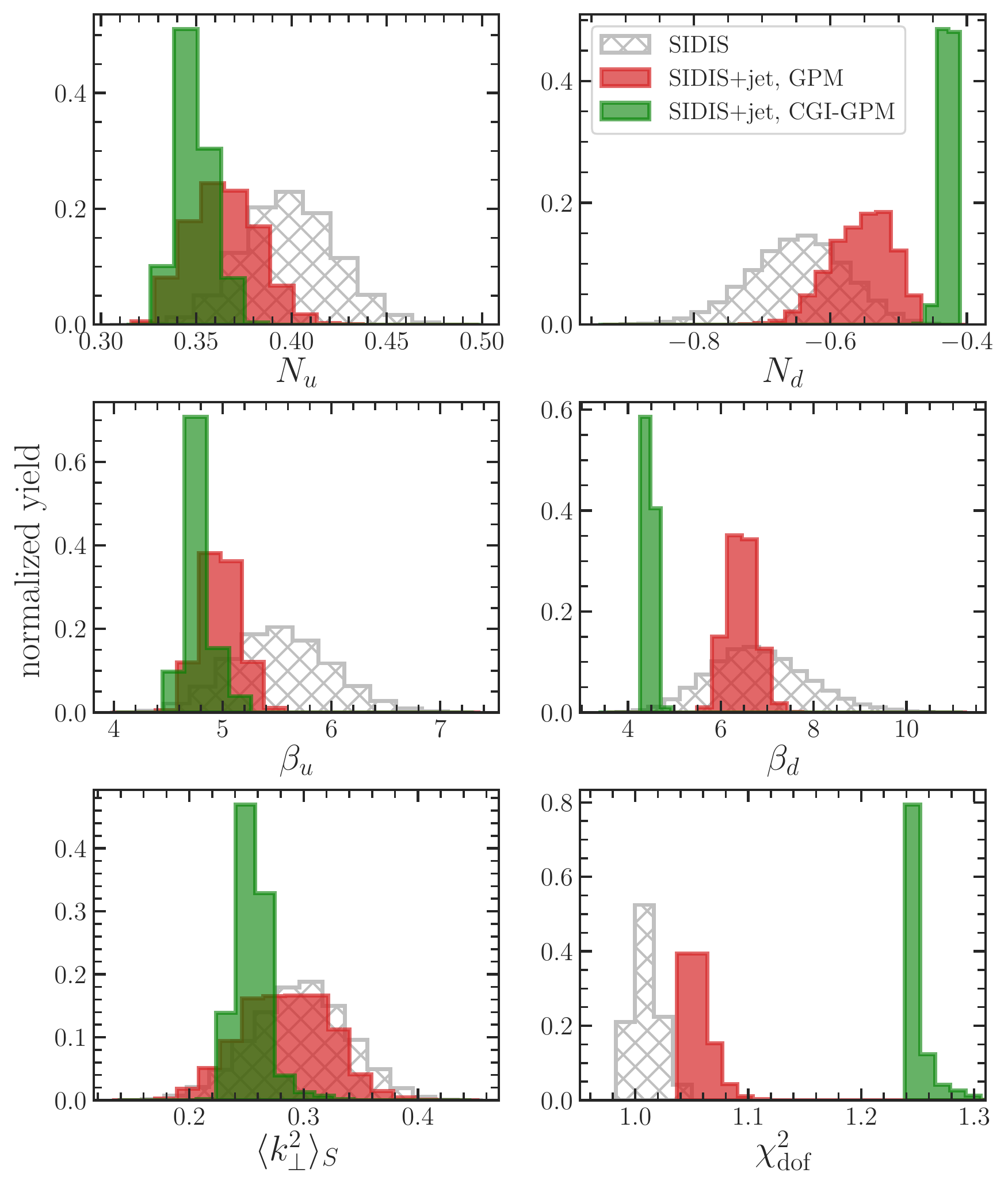}
\caption{Parameters and $\chi^2_{\rm dof}$ probability densities. Hatched histograms refer to the priors coming from SIDIS data only. Color code is the same as in Fig.~\ref{fig:reweigthing-SIDIS+jet}.}
\label{fig:pars-chi2ndof-distributions}
\end{figure}

To check the impact on the parameter and $\chi^2$ distributions, we show in Fig.~\ref{fig:pars-chi2ndof-distributions} the comparison between the priors from SIDIS and the posteriors after the reweighting. Some comments are in order: 1. the Gaussian width $\langle k^2_\perp \rangle_S$ does not vary much; 2. the $\beta_q$ parameters, governing the large-$x$ behavior of the Sivers function, change, but in a different way when applying the GPM or CGI-GPM formalisms; 3. while the normalization for the $u$-quark Sivers function $N_u$ changes slightly, $N_d$ is smaller in size in the CGI-GPM, but $f_{1T}^{\perp, d}$ is less suppressed at large $x$; 4. the $\chi^2_{\rm dof}$ after the reweighting for $N_{\rm dat}^{\rm SIDIS + jet} = 238$ slightly favors the GPM approach. For reference, we address the reader to Table I of Ref.~\cite{Boglione:2021aha}.

\begin{figure}[h!]
\centering
\includegraphics[width=.47\textwidth]{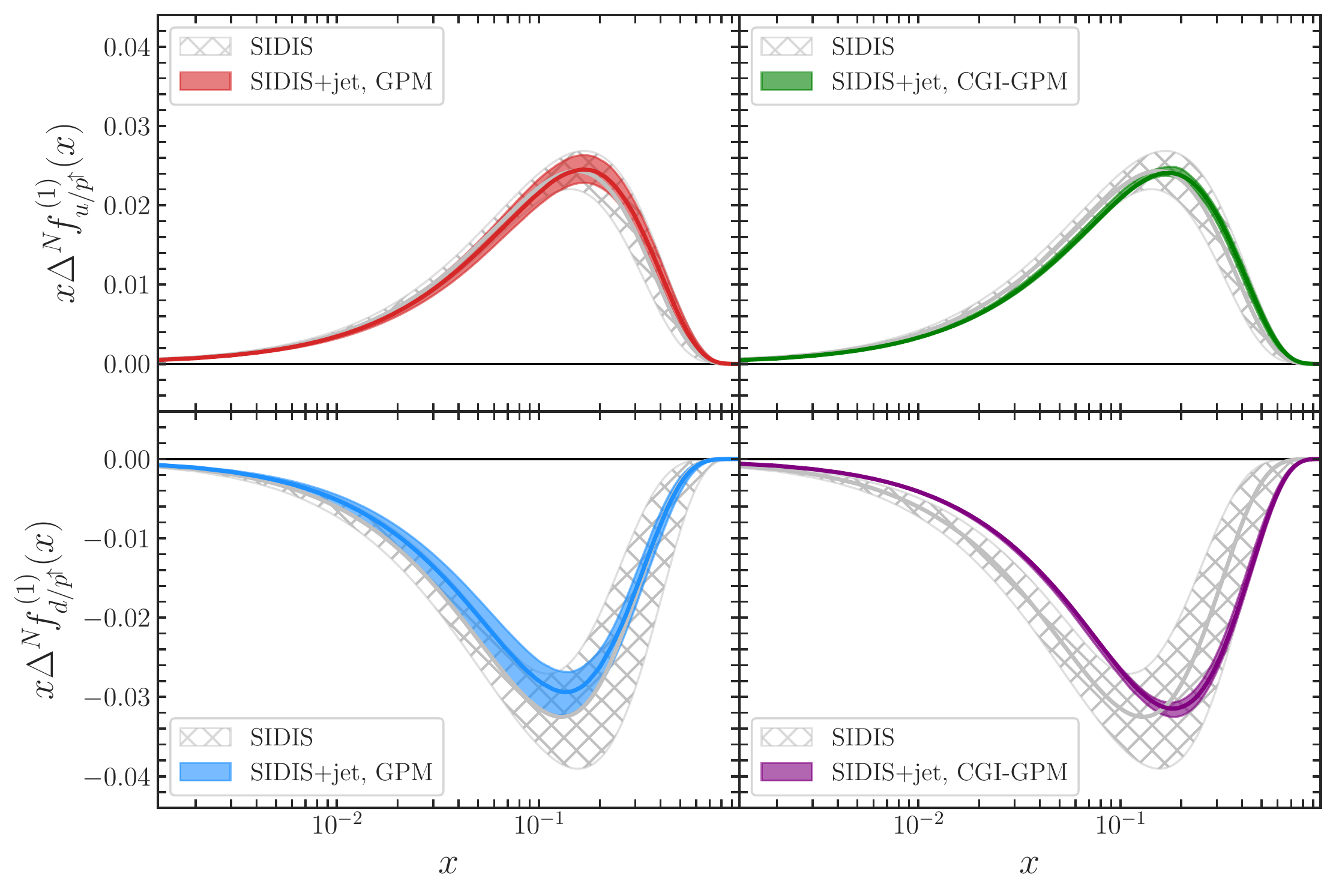}
\includegraphics[width=.45\textwidth]{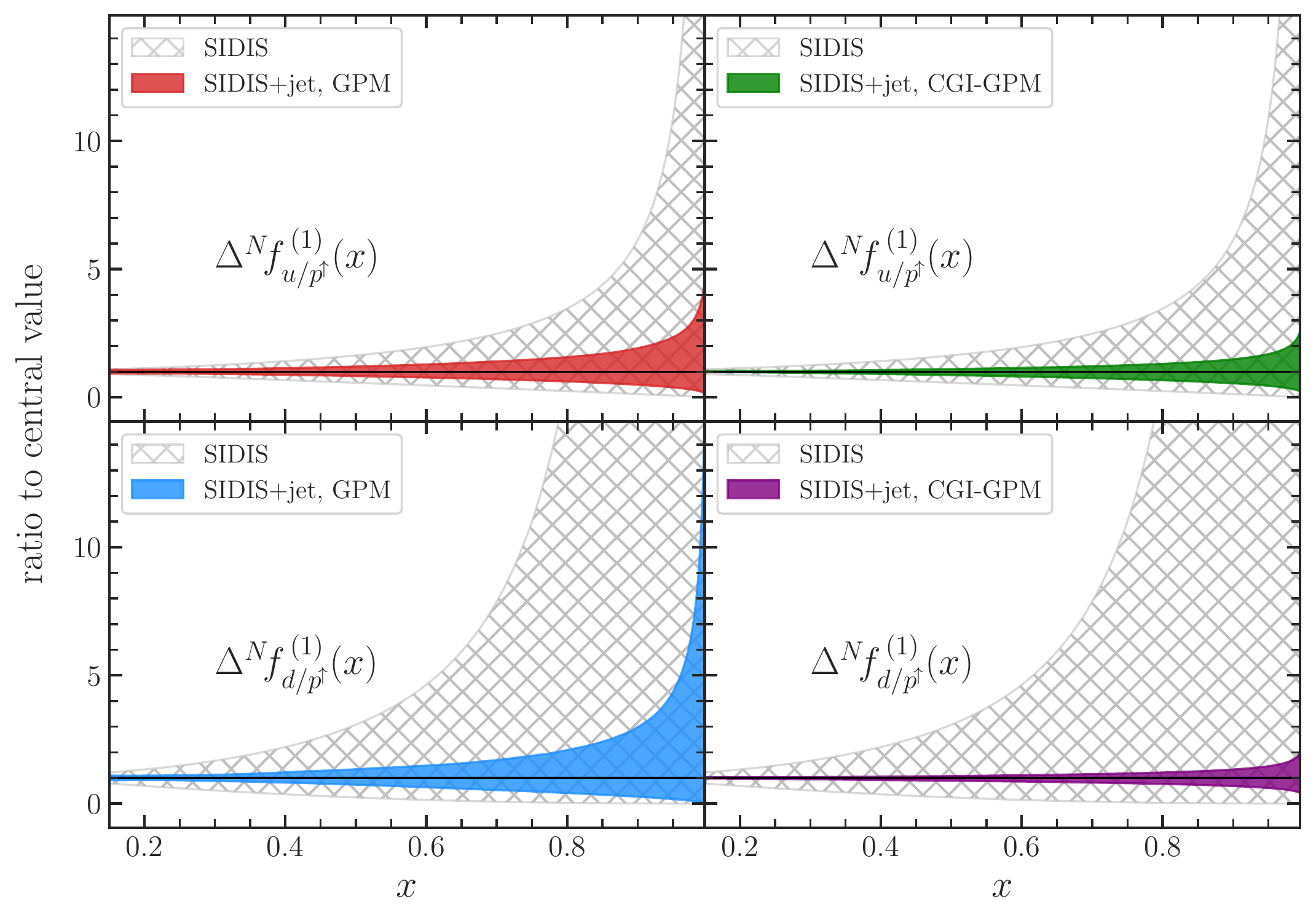}
\caption{Comparison between the Sivers first $k_\perp$-moments (left) and their values normalized to the corresponding central value (right) from SIDIS data and their reweighted SIDIS+jet counterparts in the GPM (left panels) and CGI-GPM (right panels) framework. In both plots, results for $u$- (upper panels) and $d$-quarks (lower panels) are shown. Bands correspond to a $2\sigma$ CL.}
\label{fig:Sivers}
\end{figure}

Looking now at Fig.~\ref{fig:Sivers}, one can see the impact of the reweighting procedure on the extracted functions. On the left panel, we compare the fitted first $k_\perp$-moments before and after the reweighting in the GPM and CGI-GPM approaches. The uncertainties are reduced in both cases, especially at large $x$. This appears more evident by looking at the right panel of Fig.~\ref{fig:Sivers}, where the first moments, normalized to their central values, are plotted. It is then clear that these new STAR data allows to constrain the quark Sivers function at large values of $x$, a region left unconstrained by current SIDIS measurements.

\section{Conclusion}

We have presented the first application of the Bayesian reweighting method to a TMD density, the quark Sivers function extracted from SIDIS data. Such a procedure has also been extended to the case of asymmetric uncertainties. The new STAR data allows to improve and extend the knowledge on the Sivers function at large $x$. Our findings also point to a compatibility between SIDIS and inclusive jet data.

A natural extension of this exploratory study will be a global analysis including also $A_N$ data for inclusive pion production. This would allow for a simultaneous reweighting of the Sivers, transversity and Collins functions. We expect as well that forthcoming measurements at COMPASS~\cite{Bradamante:2018ick}, JLab~\cite{Dudek:2012vr} and the future Electron Ion Collider~\cite{Boer:2011fh,Accardi:2012qut} will play a crucial role in unraveling the nucleon structure in its full complexity.

\section*{Acknowledgements}
We thank the STAR Collaboration for providing us with the experimental data~\cite{Adam:2020edg}.
We are grateful to Mauro Anselmino for his involvement in the early stages of this work.
C.F.~is thankful to the Physics Department of Cagliari University for the hospitality and support for his visit during which part of the project was done.

\paragraph{Funding information}
This project has received funding from the European Union's Horizon 2020 research and innovation programme under grant agreement STRONG-2020 - No 824093 (M.B, U.D., C.F., J.O.G.H., F.M.), by the French CNRS via the IN2P3 project GLUE@NLO and via the IEA GlueGraph (C.F.), by the P2IO Labex via the Gluodynamics project (C.F.), by the National Science Foundation under the Contract  No.~PHY-2012002 (A.P.), and by the US Department of Energy under contract No.~DE-AC05-06OR23177 (A.P.) under which JSA, LLC operates Jefferson Lab, and within the framework of the TMD Topical Collaboration (A.P.).




\begin{thebibliography}{99}

\bibitem{Sivers:1989cc}
D.~W. Sivers, {Single Spin Production Asymmetries from the Hard Scattering of
  Point-Like Constituents}, Phys.~Rev.~D 41 (1990) 83.
\newblock \href {https://doi.org/10.1103/PhysRevD.41.83}
  {\path{doi:10.1103/PhysRevD.41.83}}.

\bibitem{Sivers:1990fh}
D.~W. Sivers, {Hard scattering scaling laws for single spin production
  asymmetries}, Phys.~Rev.~D 43 (1991) 261.
\newblock \href {https://doi.org/10.1103/PhysRevD.43.261}
  {\path{doi:10.1103/PhysRevD.43.261}}.
  
\bibitem{Collins:2002kn}
J.~C. Collins, {Leading twist single transverse-spin asymmetries: Drell-Yan and
  deep inelastic scattering}, Phys.~Lett.~B 536 (2002) 43.
\newblock \href {http://arxiv.org/abs/hep-ph/0204004}
  {\path{arXiv:hep-ph/0204004}}, \href
  {https://doi.org/10.1016/S0370-2693(02)01819-1}
  {\path{doi:10.1016/S0370-2693(02)01819-1}}.

\bibitem{Brodsky:2002rv}
S.~J. Brodsky, D.~S. Hwang, I.~Schmidt, {Initial state interactions and single
  spin asymmetries in Drell-Yan processes}, Nucl.~Phys.~B 642 (2002) 344.
\newblock \href {http://arxiv.org/abs/hep-ph/0206259}
  {\path{arXiv:hep-ph/0206259}}, \href
  {https://doi.org/10.1016/S0550-3213(02)00617-X}
  {\path{doi:10.1016/S0550-3213(02)00617-X}}.
  
\bibitem{Boglione:2021aha}
M.~Boglione, U.~D'Alesio, C.~Flore, J.~O.~Gonzalez-Hernandez, F.~Murgia and A.~Prokudin,
``Reweighting the Sivers function with jet data from STAR,''
Phys. Lett. B \textbf{815} (2021), 136135
\href{https://doi.org/10.1016/j.physletb.2021.136135}{doi:10.1016/j.physletb.2021.136135}, \href{https://arxiv.org/abs/2101.03955}{arXiv:2101.03955 [hep-ph]}.

\bibitem{Gamberg:2013kla}
L.~Gamberg, Z.-B. Kang, A.~Prokudin, {Indication on the process-dependence of
  the Sivers effect}, Phys.~Rev.~Lett. 110 (2013) 232301.
\newblock \href {http://arxiv.org/abs/1302.3218} {\path{arXiv:1302.3218}},
  \href {https://doi.org/10.1103/PhysRevLett.110.232301}
  {\path{doi:10.1103/PhysRevLett.110.232301}}.
  
\bibitem{DAlesio:2004eso}
U.~D'Alesio, F.~Murgia, {Parton intrinsic motion in inclusive particle
  production: Unpolarized cross sections, single spin asymmetries and the
  Sivers effect}, Phys.~Rev.~D 70 (2004) 074009.
\newblock \href {http://arxiv.org/abs/hep-ph/0408092}
  {\path{arXiv:hep-ph/0408092}}, \href
  {https://doi.org/10.1103/PhysRevD.70.074009}
  {\path{doi:10.1103/PhysRevD.70.074009}}.

\bibitem{Anselmino:2005sh}
M.~Anselmino, M.~Boglione, U.~D'Alesio, E.~Leader, S.~Melis, F.~Murgia, {The
  general partonic structure for hadronic spin asymmetries}, Phys.~Rev.~D 73
  (2006) 014020.
\newblock \href {http://arxiv.org/abs/hep-ph/0509035}
  {\path{arXiv:hep-ph/0509035}}, \href
  {https://doi.org/10.1103/PhysRevD.73.014020}
  {\path{doi:10.1103/PhysRevD.73.014020}}.

\bibitem{DAlesio:2007bjf}
U.~D'Alesio, F.~Murgia, {Azimuthal and Single Spin Asymmetries in Hard
  Scattering Processes}, Prog.~Part.~Nucl.~Phys. 61 (2008) 394.
\newblock \href {http://arxiv.org/abs/0712.4328} {\path{arXiv:0712.4328}},
  \href {https://doi.org/10.1016/j.ppnp.2008.01.001}
  {\path{doi:10.1016/j.ppnp.2008.01.001}}.

  \bibitem{Gamberg:2010tj}
L.~Gamberg, Z.-B. Kang, {Process dependent Sivers function and implication for
  single spin asymmetry in inclusive hadron production}, Phys.~Lett.~B 696
  (2011) 109.
\newblock \href {http://arxiv.org/abs/1009.1936} {\path{arXiv:1009.1936}},
  \href {https://doi.org/10.1016/j.physletb.2010.11.066}
  {\path{doi:10.1016/j.physletb.2010.11.066}}.

\bibitem{DAlesio:2011kkm}
U.~D'Alesio, L.~Gamberg, Z.-B. Kang, F.~Murgia, C.~Pisano, {Testing the process
  dependence of the Sivers function via hadron distributions inside a jet},
  Phys.~Lett.~B 704 (2011) 637.
\newblock \href {http://arxiv.org/abs/1108.0827} {\path{arXiv:1108.0827}},
  \href {https://doi.org/10.1016/j.physletb.2011.09.067}
  {\path{doi:10.1016/j.physletb.2011.09.067}}.

\bibitem{DAlesio:2017rzj}
U.~D'Alesio, F.~Murgia, C.~Pisano, P.~Taels, {Probing the gluon Sivers function
  in $p^\uparrow p\to J/\psi\,X$ and $p^\uparrow p \to D\,X$}, Phys.~Rev.~D 96
  (2017) 036011.
\newblock \href {http://arxiv.org/abs/1705.04169} {\path{arXiv:1705.04169}},
  \href {https://doi.org/10.1103/PhysRevD.96.036011}
  {\path{doi:10.1103/PhysRevD.96.036011}}.

\bibitem{DAlesio:2018rnv}
U.~D'Alesio, C.~Flore, F.~Murgia, C.~Pisano, P.~Taels, {Unraveling the Gluon
  Sivers Function in Hadronic Collisions at RHIC}, Phys.~Rev.~D 99 (2019)
  036013.
\newblock \href {http://arxiv.org/abs/1811.02970} {\path{arXiv:1811.02970}},
  \href {https://doi.org/10.1103/PhysRevD.99.036013}
  {\path{doi:10.1103/PhysRevD.99.036013}}.

\bibitem{Adam:2020edg}
J.~Adam et~al. (STAR Collaboration), {Measurement of transverse single-spin asymmetries of $\pi^0$
  and electromagnetic jets at forward rapidity in 200 and 500 GeV transversely
  polarized proton-proton collisions}, (2020).
\newblock \href {http://arxiv.org/abs/2012.11428} {\path{arXiv:2012.11428}}.

\bibitem{Giele:1998gw}
W.~T. Giele, S.~Keller, {Implications of hadron collider observables on parton
  distribution function uncertainties}, Phys.~Rev.~D 58 (1998) 094023.
\newblock \href {http://arxiv.org/abs/hep-ph/9803393}
  {\path{arXiv:hep-ph/9803393}}, \href
  {https://doi.org/10.1103/PhysRevD.58.094023}
  {\path{doi:10.1103/PhysRevD.58.094023}}.

\bibitem{Ball:2010gb}
R.~D. Ball, V.~Bertone, F.~Cerutti, L.~Del~Debbio, S.~Forte, A.~Guffanti, J.~I.
  Latorre, J.~Rojo, M.~Ubiali, {Reweighting NNPDFs: the W lepton asymmetry},
  Nucl.~Phys.~B 849 (2011) 112, [Erratum: Nucl.Phys.B 854, 926--927 (2012),
  Erratum: Nucl.Phys.B 855, 927--928 (2012)].
\newblock \href {http://arxiv.org/abs/1012.0836} {\path{arXiv:1012.0836}},
  \href {https://doi.org/10.1016/j.nuclphysb.2011.03.017}
  {\path{doi:10.1016/j.nuclphysb.2011.03.017}}.

\bibitem{Sato:2013ika}
N.~Sato, J.~Owens, H.~Prosper, {Bayesian Reweighting for Global Fits},
  Phys.~Rev.~D 89 (2014) 114020.
\newblock \href {http://arxiv.org/abs/1310.1089} {\path{arXiv:1310.1089}},
  \href {https://doi.org/10.1103/PhysRevD.89.114020}
  {\path{doi:10.1103/PhysRevD.89.114020}}.

\bibitem{DAlesio:2020vtw}
U.~D'Alesio, C.~Flore, A.~Prokudin, {Role of the Soffer bound in determination
  of transversity and the tensor charge}, Phys.~Lett.~B 803 (2020) 135347.
\newblock \href {http://arxiv.org/abs/2001.01573} {\path{arXiv:2001.01573}},
  \href {https://doi.org/10.1016/j.physletb.2020.135347}
  {\path{doi:10.1016/j.physletb.2020.135347}}.
  
\bibitem{10.2307/2684568}
S.~Chib, E.~Greenberg, {Understanding the Metropolis-Hastings Algorithm}, The
  American Statistician 49 (1995) 327.
\newblock \href {https://doi.org/https://doi.org/10.2307/2684568}
  {\path{doi:https://doi.org/10.2307/2684568}}.  
  
  \bibitem{Boglione:2018dqd}
M.~Boglione, U.~D'Alesio, C.~Flore, J.~Gonzalez-Hernandez, {Assessing signals
  of TMD physics in SIDIS azimuthal asymmetries and in the extraction of the
  Sivers function}, JHEP 07 (2018) 148.
\newblock \href {http://arxiv.org/abs/1806.10645} {\path{arXiv:1806.10645}},
  \href {https://doi.org/10.1007/JHEP07(2018)148}
  {\path{doi:10.1007/JHEP07(2018)148}}.

  
\bibitem{Bradamante:2018ick}
F.~Bradamante, {The future SIDIS measurement on transversely polarized
  deuterons by the COMPASS Collaboration}, PoS SPIN2018 (2018) 045.
\newblock \href {http://arxiv.org/abs/1812.07281} {\path{arXiv:1812.07281}},
  \href {https://doi.org/10.22323/1.346.0045} {\path{doi:10.22323/1.346.0045}}.

\bibitem{Dudek:2012vr}
J.~Dudek, et~al., {Physics Opportunities with the 12 GeV Upgrade at Jefferson
  Lab}, Eur.~Phys.~J.~A 48 (2012) 187.
\newblock \href {http://arxiv.org/abs/1208.1244} {\path{arXiv:1208.1244}},
  \href {https://doi.org/10.1140/epja/i2012-12187-1}
  {\path{doi:10.1140/epja/i2012-12187-1}}.

\bibitem{Boer:2011fh}
D.~Boer, et~al., {Gluons and the quark sea at high energies: Distributions,
  polarization, tomography}, (2011).
\newblock \href {http://arxiv.org/abs/1108.1713} {\path{arXiv:1108.1713}}.

\bibitem{Accardi:2012qut}
A.~Accardi, et~al., {Electron Ion Collider: The Next QCD Frontier},
  Eur.~Phys.~J.~A 52 (2016) 268.
\newblock \href {http://arxiv.org/abs/1212.1701} {\path{arXiv:1212.1701}},
  \href {https://doi.org/10.1140/epja/i2016-16268-9}
  {\path{doi:10.1140/epja/i2016-16268-9}}.
  
\end{thebibliography}

\nolinenumbers

\end{document}